\documentclass[aip,graphicx]{revtex4-1}
\usepackage{graphicx}

\draft 

\begin{document}


\title{Ferroelectric Schottky diodes of CuInP$_2$S$_6$ nanosheet} 



\author{Jinyuan Yao}
\affiliation
{Department of Physics and Materials Research Institute, The Pennsylvania State University, University Park, PA 16802, USA}
\author{Yongtao Liu}
\affiliation
{Center for Nanophase Materials Sciences, Oak Ridge National Laboratory, Oak Ridge, TN 37831, USA}
\author{Shaoqing Ding}
\affiliation
{Department of Physics and Materials Research Institute, The Pennsylvania State University, University Park, PA 16802, USA}
\author{Yanglin Zhu}
\affiliation
{Nanjing Normal University, Nanjing, Jiangsu 210023,China}
\author{Zhiqiang Mao}
\affiliation
{Department of Physics and Materials Research Institute, The Pennsylvania State University, University Park, PA 16802, USA}
\author{Sergei V. Kalinin}
\affiliation
{Department of Materials Science and Engineering, The University of Tennessee, Knoxville, Tennessee 37996, USA}
\author{Ying Liu}
\email{yxl15@psu.edu}
\affiliation
{Department of Physics and Materials Research Institute, The Pennsylvania State University, University Park, PA 16802, USA}


\date{\today}

\begin{abstract}
Ferroelectricity in van der Waals (vdW) layered material has attracted a great deal of interest recently. CuInP$_2$S$_6$ (CIPS), the only vdW layered material whose ferroelectricity in the bulk was demonstrated by direct polarization measurements, was shown to remain ferroelectric down to a thickness of a few nanometers. However, its ferroelectric properties have just started to be explored in the context of potential device applications. We report here the preparation and measurements of metal-ferroelectric semiconductor-metal heterostructures using nanosheets of CIPS obtained by mechanical exfoliation. Four bias voltage and polarization dependent resistive states were observed in the current-voltage characteristics, which we attribute to the formation of ferroelectric Schottky diode, along with switching behavior. 
\end{abstract}


\maketitle 


Ferroelectric semiconductors can have interesting properties that are potentially useful for microelectronic applications\cite{Si_AlphaIn2Se3FeSmFET_2019}. For example, when a ferroelectric semiconductor is in a close electrical contact with a metal, ferroelectric polarization can have a strong effect on the Schottky barrier and the depletion near the interface, leading to rectifying behavior and ferroelectric Schottky diodes (FSDs) \cite{Wang_BFOSwitchableDiodeEffect_2011, Blom_PTOFSD_1994}. So far FSDs have been studied using mostly an oxide as the ferroelectric layer, including $\mathrm{PbTiO}_3$ \cite{Blom_PTOFSD_1994}, $\mathrm{BiFeO}_3$ \cite{Wang_BFOSwitchableDiodeEffect_2011, Matsuo_BFOSDE_2015}, Pb(Zr$_{0.2}$Ti$_{0.8}$)O$_3$ \cite{Pintilie_PZTFSD_2010}, $\mathrm{Zn}_x\mathrm{Cd}_{1-x}\mathrm{S}$ \cite{Sluis_ZnCdSFSD_2003}. The polarization was found to control the resistive state of FSD, which may be useful for nonvolatile memory applications. Comparing with the traditional one-transistor-one-capacitor ferroelectric random access memory (1T1C FeRAM) whose read/write process is destructive, FSDs have the potential to achieve improved endurance due to its nondestructive readout process \cite{Eshita_FRAM_2014}.

\begin{figure}
\includegraphics[width=250pt]{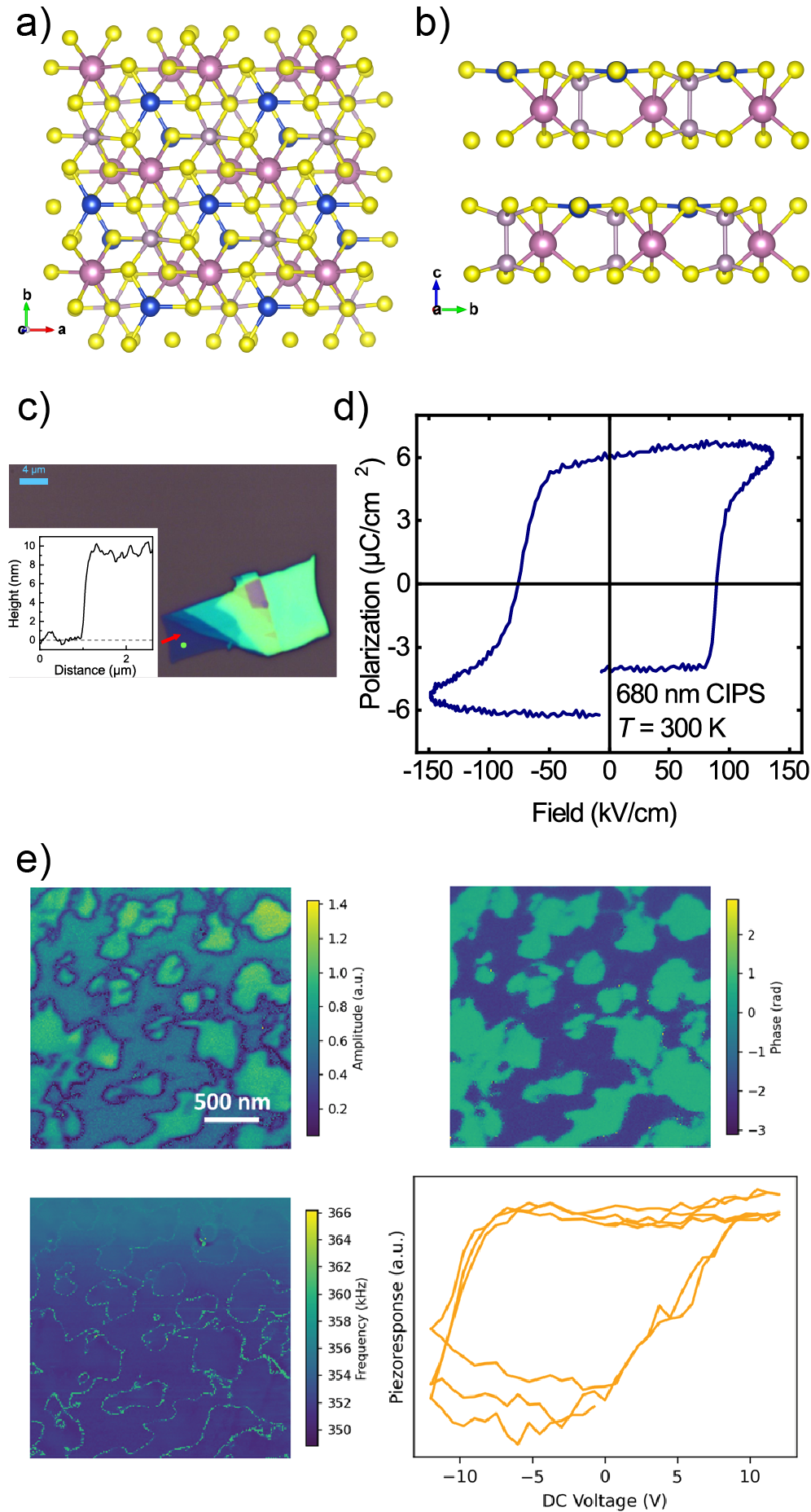}
\caption{\textbf{Structure and material characterization of CIPS.}  a-b) Top and side views of the crystal structure of CuInP$_2$S$_6$. c) Atomic force microscopy (AFM) of an exfoliated nanosheet of CIPS. A line scan is shown in the inset. d) Hysteresis loop from a 680-nm-thick crystal of CIPS measured at room temperature. e) Piezoforce force microscopy (PFM) characterization of a thin crystal of CIPS on top of bottom electrode formed by 5 nm thick Ti and 45 nm Au films deposited on a Si substrate with thermally grown SiO$_2$ surface.\label{Fig1}}
\end{figure}

 Van der Waals (vdW) layered ferroelectric semiconductors that can be exfoliated into nanosheets are attractive for the study of FSD as an atomically sharp interface with a metal can be obtained. ${\mathrm{CuInP}}_{2}{\mathrm{S}}_{6}$ (CIPS) features a layered monoclinic crystalline structure (Fig. \ref{Fig1}a, \ref{Fig1}b) with a $c$-axis lattice constant roughly 1.36 nm \cite{Maisonneuve_CIPSStructure_1997}. The bulk ferroelectric transition temperature ($T_c$) was found to be around 315~K, with a remnant polarization of 2.55~${\mathrm{\mu C/cm}^{2}}$ and a coercive field of 77~ kV/cm\cite{Simon_CIPSBulkFerroelectricity_1994}. More recent measurements suggest that the bulk remnant polarization can be as high as 4.5 ${\mathrm{\mu C/cm}^{2}}$ with a coercive field as low as 30~kV/cm \cite{Zhou_CIPSPolarization_2020}. Piezoelectric force microscopy (PFM) suggests that piezoelectricity and ferroelectricity persists down to a 4 nm thickness \cite{Liu_CIPS4nm_2016, Io_CIPSPFM_2022} even though its remnant polarization cannot be measured due to leakage current. As-grown crystals of CIPS are typically hole-doped semiconductors \cite{Samulionis_CIPSSemiconductor_2001, Yu_CIPSCarrierTypeExperiment_2021} with a band gap of 2.8 eV \cite{Zhou_CIPSReview_2021}.
 
 Switchable behavior was found previously in metal-CIPS-metal structures\cite{Li_CIPSSynapse_2020, Si_CIPSFeFET_2018, Jiang_CIPSIonMirgation_2022, Kong_CIPSPhotoDep_2022, Chen_CIPSNeuro_2022}. However, the connection between the observed switchable behavior and the switching of the polarization in the ferroelectric layer was not well established. A potential issue is that the existence of vacancies and charge traps at the interfaces may also lead to switching behavior \cite{Ding_BZOSROSwitchableDiode_2016}. In the current study we prepared Au/CIPS/Cr and Au/CIPS/Ni metal-ferroelectric-metal heterostructures and demonstrated resistive switchings with an ON/OFF ratio up to 10$^3$, as shown previously in the studies of BaZrO$_3$/SrRuO$_3$ structures\cite{Ding_BZOSROSwitchableDiode_2016}  and Sr doped BiFeO$_3$ thin films \cite{Guo_BFSOSwitchableDiode_2013}. The issue of vacancy and Cu ion motion $vs.$ polarization switching was explored by examining behaviors of devices of different thickness and electrodes.

 Thin and ultrathin crystals of CIPS were obtained from single crystals grown by chemical vapor transport (CVT) \cite{Niu_CIPSCVT_2019}. To grow single crystals of CIPS, Cu, In, P, and S powder were mixed in a stoichiometric ratio and loaded into a quartz tube (10 mm in diameter, 18 cm length). The transport agent, I$_2$, was added. The quartz tube was sealed under vacuum and then heated up in a horizontal double-zone furnace with the hot and cold ends set at 750 °C and 650 °C, respectively. After 14 days, the furnace was shut down, and the quartz tube was naturally cooled down to room temperature. The yellow plate-like single crystals were found at the cold end of the quartz tube. CIPS crystals were mechanically exfoliated onto polydimethylsiloxane (PDMS) and then transferred onto the bottom electrode prepared by depositing 5 nm thick Ti followed by 45 nm thick Au on a Si/SiO$_2$ substrate. The thickness of cleaved crystal was estimated by its color and faintness under the optical microscope and determined more precisely using atomic force microscopy (AFM) (Bruker Dimension). Top electrodes were made by e-beam deposition of Ni (20 nm thick) and Au (30 nm thick) or Cr (20 nm thick) and Au (30 nm thick) sequentially followed by maskless photolithography. Raman measurements were conducted on Horiba LabRAM HR. DC Current-voltage measurements were conducted on probe station Form Factor 11000B with Keithley 4200-SCS Parameter Analyzer.

In Fig. 1d we show a hysteresis loop of a 680-nm-thick CIPS crystal obtained by mechanical exfoliation and polarization $vs.$ electric field (P(E)). For this structure, the bottom and top electrodes (BE and TE) are Au and Pt, respectively. Thinner crystals of CIPS can also be obtained as shown in Fig. 1c. At such thickness the polarization is difficult to measure because the leakage currents are usually very large. The remnant polarization $P_r$ and coercive field $E_c$ in the bulk are 2.55 ${\mathrm{\mu C/cm}^{2}}$ and 77 kV/cm, respectively \cite{Maisonneuve_CIPSStructure_1997}. For our thick crystals, the $P_r$ and $E_c$ were found to be 6 ${\mathrm{\mu C/cm}^{2}}$ and 89 kV/cm, respectively. The open hysteresis loop seen in Fig. 1d indicates that the polarization state did not recover its initial state after the electrical sweep due to incomplete switching of domains or the presence of defects \cite{Zhao_CIPSTempEffect_2020}.

\begin{figure}
\includegraphics[width=300pt]{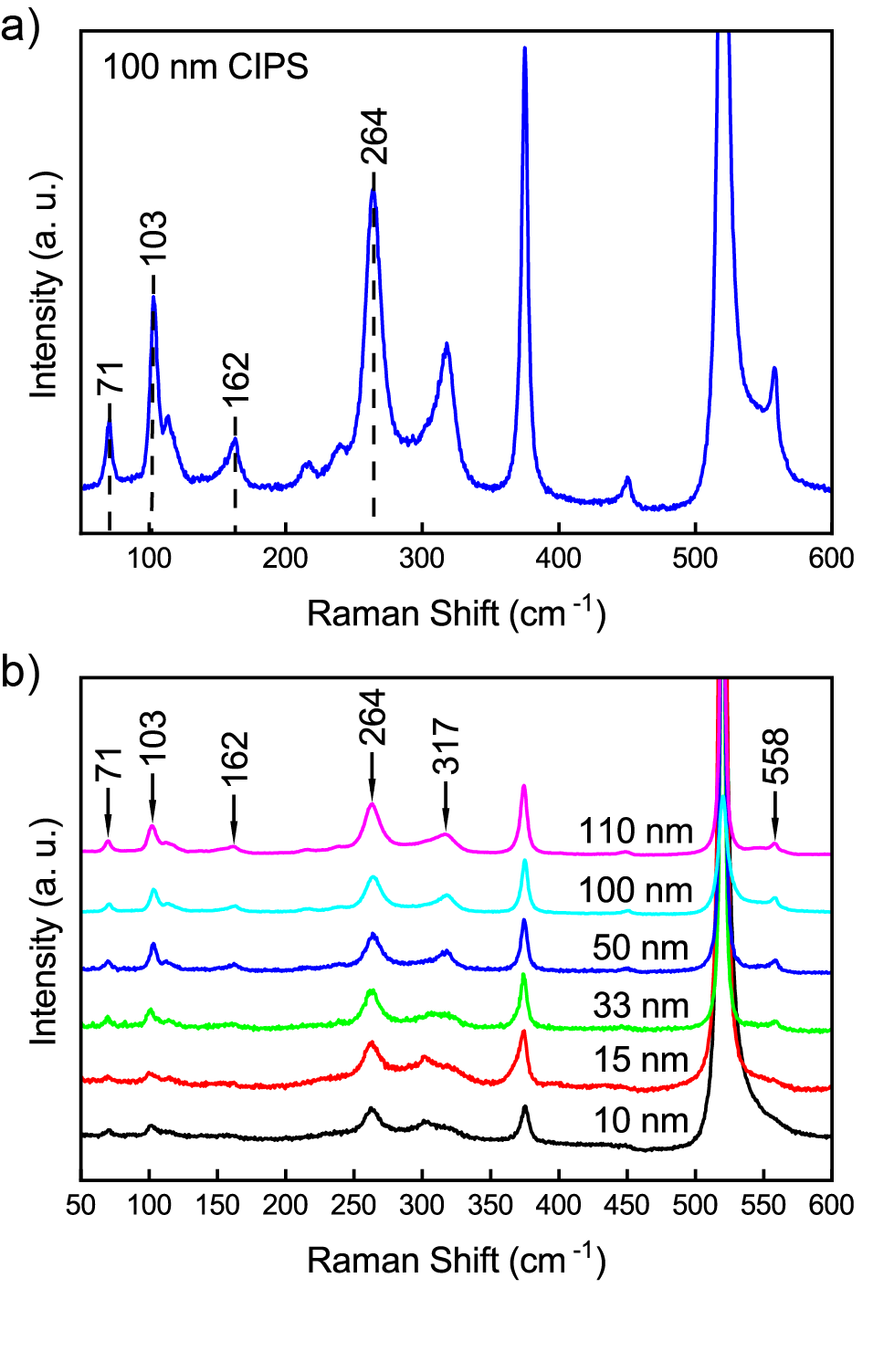}
\caption{\textbf{Raman spectra of thin CIPS crystals.} a) Raman spectrum of a 100-nm-thick CIPS crystal. Four peaks that only occur in the ferroelectric phase are labeled. b) Thickness dependence of Raman spectra of CIPS. Peaks that change significantly as thickness decreases are labeled.\label{Fig2}}
\end{figure}

We carried out Raman spectroscopy measurements of thin crystals of CIPS as its thickness is reduced. Figure 2a shows the Raman spectrum of a 100-nm-thick CIPS flake. Four peaks that are only present in ferroelectric phase (71, 103, 162 and 264 ${\mathrm{cm}^{-1}}$) \cite{Neal_DavidVanderbilt_CIPSRaman_2022} were observed. The peak at 71 ${\mathrm{cm}^{-1}}$ is related to the out-of-plane motion of Cu ions, which is believed to be responsible to the ferroelectricity in CIPS\cite{Zhou_CIPSReview_2021}. Other peaks correspond to the out-of-plane motion of P-P dimers and in-plane motion of S atoms. Figure 2b shows the Raman spectra from thin and ultrathin crystals of CIPS with different thicknesses ranging from 10 nm to 110 nm. All four peaks mentioned above are seen with their positions essentially the same as the thickness is reduced, suggesting that the ferroelectricity is robust down to 10 nm. However, significant changes are seen in two peaks found at 317 and 558 ${\mathrm{cm}^{-1}}$, respectively, which are not changed significantly until the thickness becomes smaller than 33 nm. As the thickness decreases further, the 317 ${\mathrm{cm}^{-1}}$ shifts to lower wave number while the width of this peak increases. The peak at 558 ${\mathrm{cm}^{-1}}$ does not shift but the intensity of this peak decreases as the thickness goes down. For 15- and 10-nm-thick crystals, this peak essentially vanishes. The implications of these observations are to be understood.

\begin{figure}
\includegraphics[width=300pt]{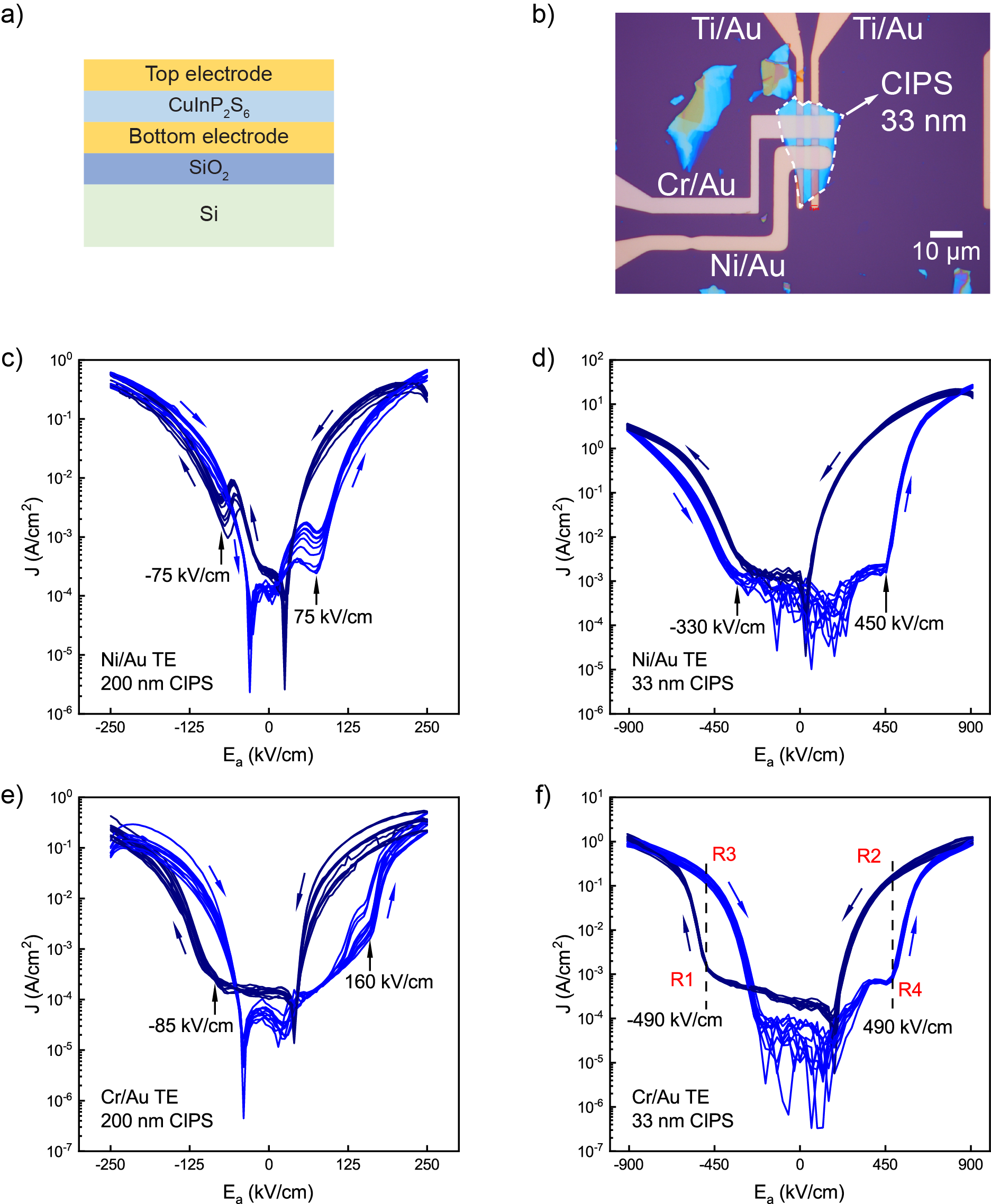}
\caption{\textbf{Current-voltage characteristics and resistive states.} (a) Schematic of our device structure. (b) Optical microscope images of two trilayer devices prepared on a 33~nm-thick crystal of CIPS with a top electrode of Cr and Ni, respectively, both of which were topped by a 30 nm thick Au layer. Other two trilayer devices were also prepared similarly on a 200~nm-thick crystal of CIPS (Fig. S1a). (c-f) Current-voltage characteristics measured from four trilayer devices possessing either a Ni/Au or Cr/Au top TE but different thicknesses of CIPS crystals (200 and 30~nm, respectively) as indicated. Eleven hysteresis loops were recorded. Negative voltages for negative currents were converted to be positive to facilitate semi-log plotting. Positive-to-negative and negative-to-positive branches are labelled by different colors. The voltages near which a jump in current density occurs are indicated. Four resistance states (corresponding to Fig. 4(c)-(f)) are labelled in (f).\label{Fig3}}
\end{figure}

In Fig. \ref{Fig3}, the curves of current density ($J$) $vs$. applied electric field ($E_a$) ($J$-$E_a$) are shown for four heterostructures prepared on 200- and 33-nm-thick crystals of CIPS, with BEs of all four devices Au (with a Ti underlay) and the TEs Cr and Ni (with a Au overlay), respectively. Hysteresis loops were observed in all four devices. For the two devices with Ni as the TE, two deep minima are seen at two different non-zero values of $E_a$. When the magnitude of $E_a$ increases for both the positive and negative biases, the Ni-200nm device was found to show an initial increase in $J$, a drop, and then a rise sharply again at higher $E_a$ values starting around 75 kV/cm, which is roughly the reported coercive fields of bulk CIPS \cite{Zhao_CIPSTempEffect_2020, Li_CIPSSynapse_2020}. For the Ni-33nm device, the corresponding values for the sharp rise in $J$ are 450 kV/cm on the positive and -330 kV/cm on the negative side of $E_a$, respectively, also consistent with increased coercive field of thin crystals of CIPS. Therefore, the rapid increase in $J$ appears to be associated with the switching of the ferroelectric polarization. Indeed, the butterfly-shaped hysteresis in the $J$-$E_a$ curves of these two heterostructures are qualitatively consistent with the ferroelectric switching. 

However, complex behavior was also found in devices featuring a Ni TE. For the Ni-200nm device, the drop of $J$ after reaching two secondary peaks as described above suggests the depletion of mobile charges even as the magnitude of $E_a$ was increasing. Moreover, the ratio of upper branch to lower branch of the hysteresis loop on the negative bias side is much smaller than the ratio on the positive bias side. For the Ni-33nm device, the hysteresis loop on the negative bias side actually has a different direction than that of the Ni-200nm device, as seen in Fig. \ref{Fig3}c and d.     

For devices with a Cr TE (Fig. \ref{Fig3}e and f), the behavior seems to be simpler than those with Ni as TE. The secondary peaks seen in $J$-$E_a$ curves of the Ni-200nm device are absent in both the positive and negative voltage sweeps. Interestingly, the $J$-$E_a$ curves are seen to stay relatively flat at the low bias voltages, making the negative-to-positive sweeps symmetric without showing rectification. In addition, $J$ values in the $J$-$E_a$ curves for the positive to zero sweep are small and the hysteresis loop in $J$-$E_a$ curve becomes counterclockwise for both devices.

\begin{figure}
\includegraphics[width=300pt]{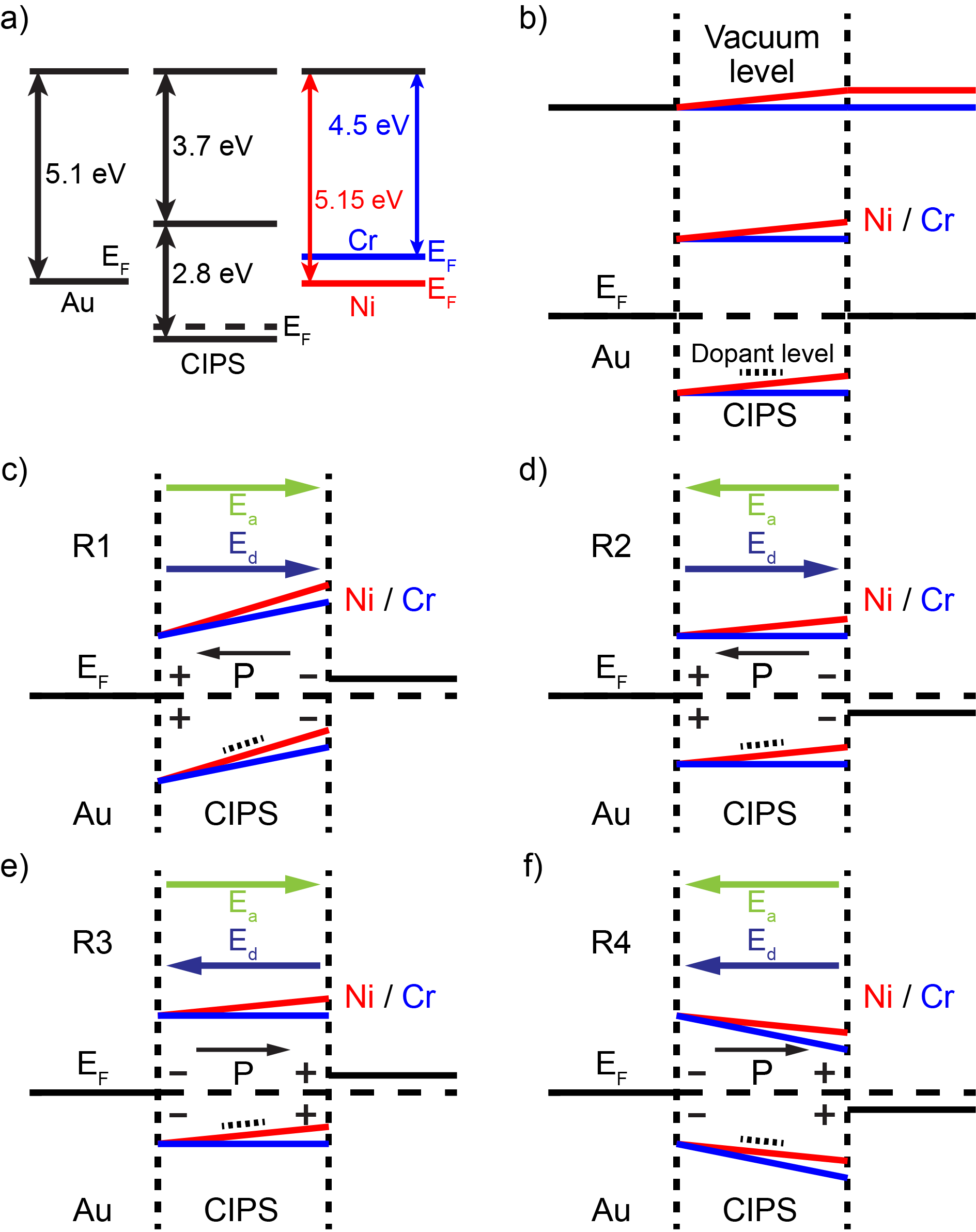}
\caption{\textbf{Band diagrams for Au/CIPS/Cr and Au/CIPS/Ni structures.} Band diagrams corresponding to a) down pointing polarization and negative bias, b) down pointing polarization and positive bias, c) up pointing polarization and negative bias, and d) up pointing polarization and positive bias.\label{Fig4}}
\end{figure}

Similar $J$-$E_a$ characteristics were seen previously in BiFeO$_3$ \cite{Lee_BFODefectMediated_2016}, which were attributed to the defects formed during the synthesis. The important question is whether the switching behavior seen in the present work can also be related to defects or ionic motion in CIPS. Indeed, sulfur vacancies is common in CIPS due to its volatility, resulting in the 2p peak shift seen previously in the X-ray photo spectroscopy (XPS) studies \cite{Yu_CIPSCarrierTypeExperiment_2021}. In addition, it has been widely reported \cite{Zhang_CIPSAnisotropicIonMigr_2021, Jiang_CIPSIonMirgation_2022, Neumayer_CIPSIonicCap_2022} that mobile Cu ions are present in CIPS, which would in principle also affect the conductivity of our devices. However, it is unlikely that the motion of S vacancies and/or Cu ions is plays an important role in our devices. First, the voltage was swept at around 1.3 V/s, which is so slow that the hysteresis from Cu ion migration should only appear on one side of the voltage sweeping\cite{Jiang_CIPSIonMirgation_2022}, opposite from what was seen experimentally . In addition, the motion of the S vacancies and/or Cu ions is unlikely to be affected by the electrodes as they reside mostly in the interior of the CIPS crystals, which is also different from what was seen experimentally. We propose that the formation of a Schottky barrier in a ferroelectric semiconductor and the effect of polarization on the barrier height is responsible to the characteristics of our devices. The relative positions of vacuum levels, Fermi levels and conduction band minima of metal electrodes and CIPS are shown in Fig. 4a. The depletion width without bias can be estimated by\cite{Neamen_SmPhysicsDevices_2012}

\begin{equation}
    W=\sqrt{\frac{2 \epsilon \epsilon_0 V_{bi}}{q N_A}},
\end{equation}

where $\epsilon$, $V_{bi}$, $q$ and $N_A$ represent dielectric constant of CIPS, built-in potential, elementary charge and density of the acceptors. According to a previous report, dielectric constant of CIPS is 40 \cite{Zhou_CIPSReview_2021}. When $V_{bi}=1.5 \mathrm{V}$ and $N_A=10^{18} \mathrm{cm^{-3}}$, depletion width $W$ is around 80 nm, much larger than 33 nm. Thus, 33 nm CIPS should be fully depleted.

A Schottky barrier is formed with band bending expected only in thick CIPS but not in thin ones as stated above. When the polarization is present in CIPS, the negative bound charge tends to suppress the Schottky barrier, while the positive bound charge will increase the Schottky barrier height at the interface (Figs. S4b and c). The device can be modeled as a diode with the Schottky barrier height controlled by the ferroelectric polarization. In devices prepared on the thin crystal, the CIPS is totally depleted so the band bending cannot form (Fig. 4b). The heights of the Schottky barriers on the two interfaces and the sample resistance are affected by the polarization. When the polarization points down (towards Au electrode), the barrier height on Cr-CIPS or Ni-CIPS interface is larger without bias voltage. When negative bias voltage is applied (Figs. 4c and e), barrier height for electron transport is lowered, leading to a low resistance state. Positive bias voltage would then correspond to a high resistance state (Fig. 4d). When polarization points up, in Cr-33nm device, Au-CIPS interface has a larger barrier height. Thus, the device is in low resistance state when positive bias voltage is applied (Fig. 4f). For Cr-33nm device, the device behaves as a diode that is forward biased when the applied field is parallel with depolarization field (Figs. 4c and f), leading to the butterfly shape $J$-$E_a$ curves in Fig. 3f. The difference between the hysteresis directions of Cr-33 nm and Ni-33 nm devices may be attributed to the different work functions of Cr and Ni. Larger work function of Ni leads to a larger barrier height on Ni-CIPS interface. When polarization points up, different from Cr-33 nm device, the barrier height of Ni-CIPS interface is around similar as the one of Au-CIPS interface without bias voltage, making it the same for positive and negative bias voltages (Figs. 4e and f) in Ni-33nm sample, leading to the symmetric, relatively flat $J$-$E_a$ curve at low bias voltage (Fig. 3d).

Combination of positive/negative bias and two ferroelectric polarization states would give four resistance states. For Cr-33 nm device, resistance values of the two low (high) resistive states under two ferroelectric states are close, possibly due to the coincidence of the Schottky barrier height of Au-CIPS and Cr-CIPS contact. Even though Au and Cr have very different work function values, Au-CIPS interface was formed by van der Waals transfer while Cr-CIPS interface formed during evaporation. When the Cr is replaced by other materials with very different work functions, such as Ni, this coincidence should not be present. Indeed, the Ni-33 nm device is seen to show four distinct resistive states.

Previous research showed that CIPS has two extra ferroelectric states with larger polarization (high polarization states) \cite{Brehm_CIPSQuadruple_2020}. In principle, the peaks occurred in Fig. 3e can correspond to the switching from low polarization to high polarization state. However, the potential barrier between high and low polarization is quite small \cite{Zhou_CIPSPolarization_2020}, making high polarization states metastable. It is difficult to observe the transition to high polarization states by transport measurements. Moreover, Cr-200nm sample did not show peaks in its $J$-$E_a$ curve, suggesting that observed peaks are not associated with high polarization states.

For devices of thick CIPS, the $J$-$E_a$ curves show butterfly-shape hysteresis loops. For Ni-200nm device, two secondary peaks in the magnitude of $J$ are seen when the magnitude of $E_a$ was swept from low to high values in both directions. As seen in Figs. S4e and g, the band bending leads to a "potential well" for mobile electrons migrated from TE near the CIPS/TE interface. Because of the work function difference, the "potential well" near the CIPS/TE interface for the TE of Ni is more shallow than that for Cr, which means that these electrons can be removed from the "potential well", leading to a peak in $J$. However, after these electrons trapped in the "potential" well are all removed, $J$ will start to drop, leading to a negative conductance as the magnitude of $E_a$ increases further.

With respect to the resistance ratio, for devices with Cr as the TE, the resistance ratios under positive and negative biases are similar. In Cr-200nm and Cr-33nm devices, a resistance ratio $\sim$ 100 is reached. For devices with Ni as top electrodes, the resistance ratio under negative bias is much smaller than the ratio under positive bias. Resistance ratios under positive and negative bias are $\sim$ 100 and 10, respectively, for Ni-200nm devices. The largest resistance ratio $\sim$ 1800 was obtained in Ni-33nm device under positive bias (under applied field $\sim$ 450 kV/cm, labelled in Fig. 3f). Same as Ni-200nm device, resistance ratio under negative bias is much smaller, only $\sim$ 6 (under applied field $\sim$ -450 kV/cm, labelled in Fig. 3d). Four resistance states ranging from M$\Omega$ to G$\Omega$ were achieved in Ni-33nm.

In conclusion, we have prepared Au/CIPS/Cr and Au/CIPS/Ni heterostructures using nanosheets of CIPS obtained by mechanical exfoliation. We found four resistive states which are determined by bias and polarization as well as switching behavior in the $J$-$E_a$ characteristics, which we attribute to the formation of FSD. We also found that thicknesses of the ferroelectric semiconductor and the work functions of the electrodes play an important role for these devices. The performance of these FSDs may be optimized, making them potentially useful for non-volatile memory applications.

\section*{SUPPLEMENTARY MATERIAL}
See supplementary material for optical images of Cr- and Ni-200 nm devices, extra I-V characteristics and band diagrams for devices with thick films.


%
%

%

\section*{ACKNOWLEDGMENTS}
The growth of single crystals of CIPS used in this work was supported by the National Science Foundation through the Penn State 2D Crystal Consortium-Materials Innovation Platform (2DCC-MIP) under NSF cooperative agreement DMR-1539916, and DMR-2039351. Bulk of the work, including device, characterization, and data analysis was supported as part of the center for 3D Ferroelectric Microelectronics (3DFeM), an Energy Frontier Research Center funded by the U.S. Department of Energy (DOE), Office of Science, Basic Energy Sciences under Award Number DE-SC0021118. Useful discussions with Profs. S. Trolier-Mckinstry and T. Jackson are gratefully acknowledged.

\section*{AUTHOR DECLARATIONS}
\subsection*{Conflict of Interest}
The authors have no conflicts to disclose.
\subsection*{Author Contributions}
\textbf{Jinyuan Yao}: Conceptualization (supporting); Data curation (lead); Formal analysis (lead); Writing – original draft (lead); Writing – review and editing (supporting). \textbf{Yongtao Liu}: Data curation (supporting); validation (supporting). \textbf{Shaoqing Ding}: Data curation (supporting); validation (supporting). \textbf{Yanglin Zhu}: Resources (supporting). \textbf{Zhiqing Mao}: Funding Acquisition (supporting); Resources (supporting); validation (supporting). \textbf{Sergei Kalinin}: Resources (supporting). \textbf{Ying Liu}: Conceptualization (lead); Funding Acquisition (lead); Resources (lead); validation (supporting); Writing – review and editing (lead).

\section*{DATA AVAILABILITY}
The data that support the findings of this study are available from the corresponding author upon reasonable request.

\bibliography{CIPS}

\end{document}



\title{Supplementary Material \\ Ferroelectric Schottky diodes of CuInP$_2$S$_6$ nanosheet}

\author{Jinyuan Yao}
\affiliation
{Department of Physics and Materials Research Institute, The Pennsylvania State University, University Park, PA 16802, USA}
\author{Yongtao Liu}
\affiliation
{Center for Nanophase Materials Sciences, Oak Ridge National Laboratory, Oak Ridge, TN 37831, USA}
\author{Shaoqing Ding}
\affiliation
{Department of Physics and Materials Research Institute, The Pennsylvania State University, University Park, PA 16802, USA}
\author{Yanglin Zhu}
\affiliation
{Nanjing Normal University, Nanjing, Jiangsu 210023,China}
\author{Zhiqiang Mao}
\affiliation
{Department of Physics and Materials Research Institute, The Pennsylvania State University, University Park, PA 16802, USA}
\author{Sergei V. Kalinin}
\affiliation
{Department of Materials Science and Engineering, The University of Tennessee, Knoxville, Tennessee 37996, USA}
\author{Ying Liu}
\affiliation
{Department of Physics and Materials Research Institute, The Pennsylvania State University, University Park, PA 16802, USA}
\email{yxl15@psu.edu}

\date{\today}

\maketitle

\begin{figure}
    \includegraphics[width=300pt]{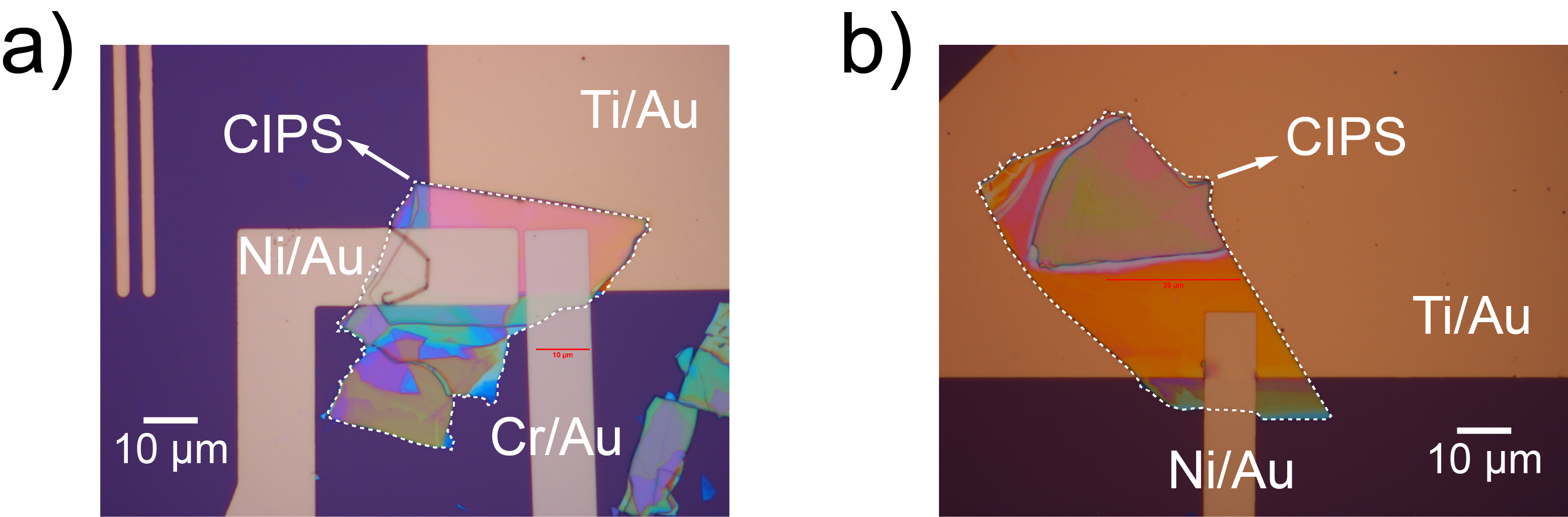}
    \caption*{FIG. S1. Microscopic pictures of devices prepared on 200 nm CIPS (a) and another Au/CIPS/Ni structure device prepared on a thick CIPS flake. The thickness of the CIPS flake in (b) is estimated to be few hundreds of nanometers based on the color.\label{FigS1}}   
\end{figure}

\begin{figure}
    \includegraphics[width=300pt]{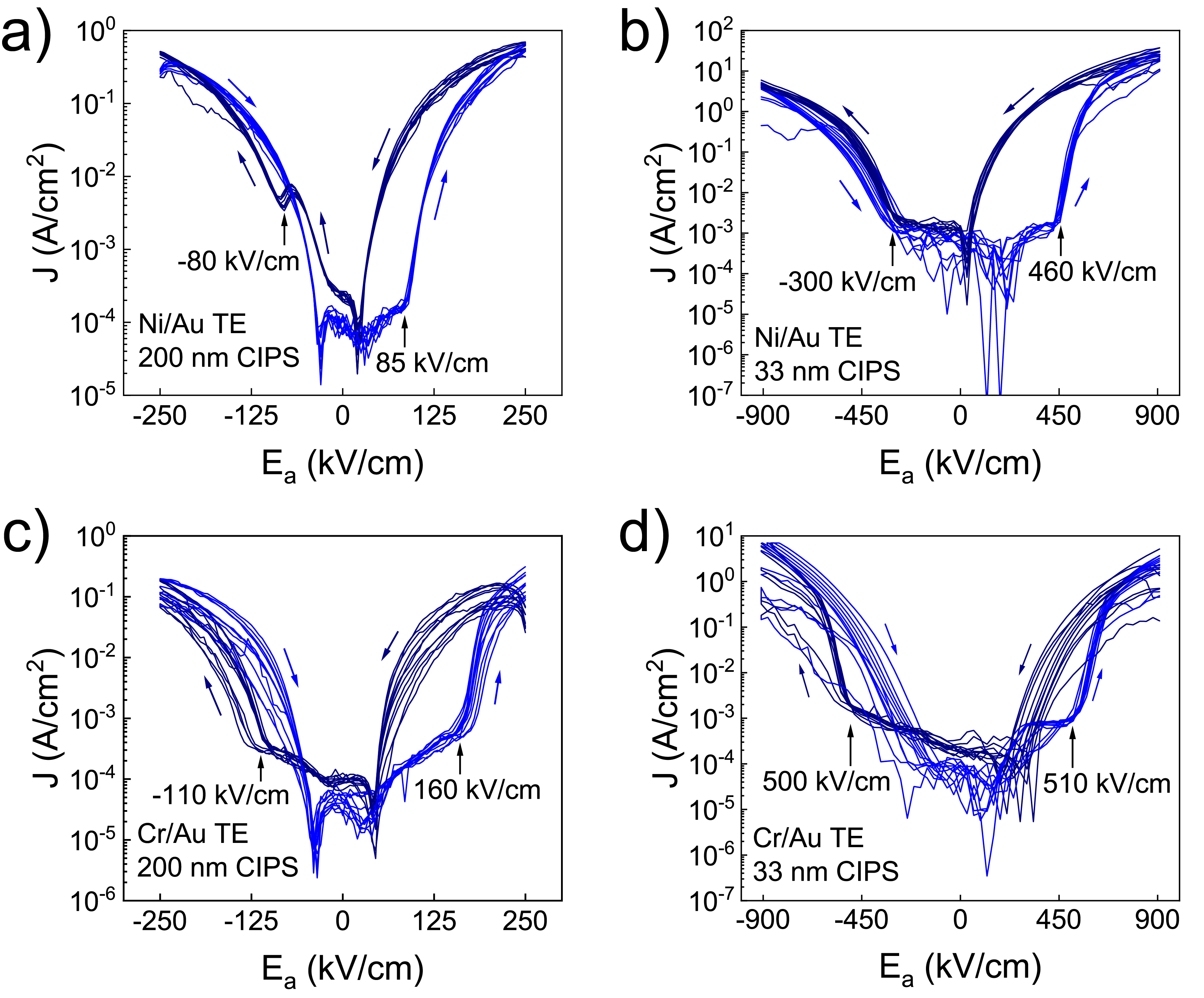}
    \caption*{FIG. S2. The J-E$_a$ curves from the second set of measurements taken on Ni-200nm (a), Ni-33nm (b), Cr-200nm (c) and Cr-33nm (d) devices. Around 10 cycles of sweep are shown here. Degradation is found in all devices, especially in devices using Cr as top electrodes.\label{FigS2}}
\end{figure}

\begin{figure}
    \includegraphics[width=450pt]{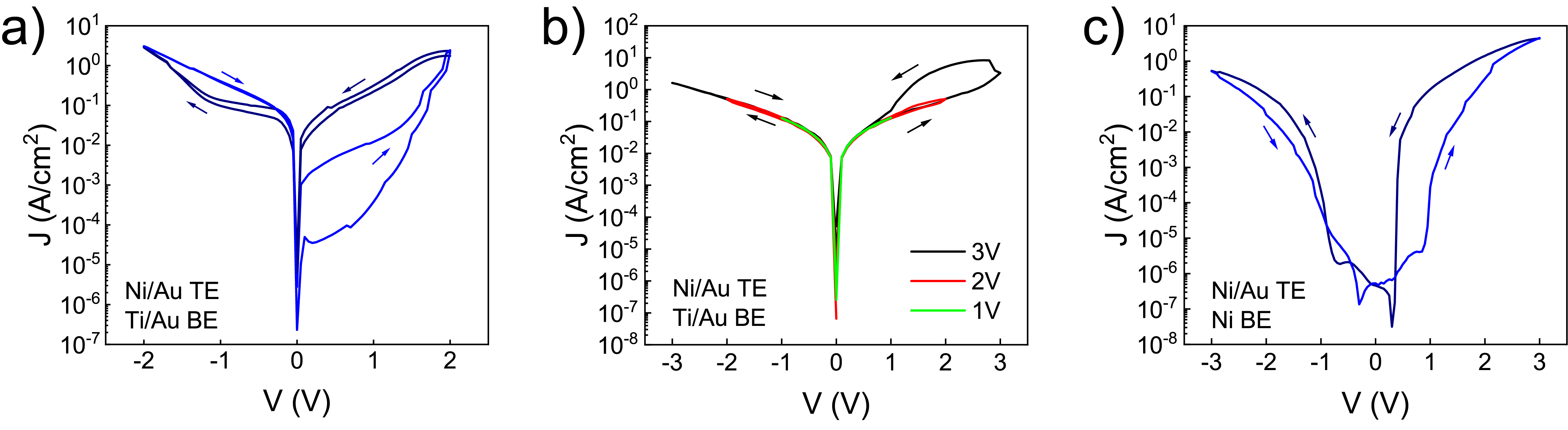}
    \caption*{FIG. S3. I-V characteristics of other samples. (a) I-V curves of Au/CIPS/Ni structure. Two curves were measured from two different Ni top electrodes. (b) I-V curves of Au/CIPS/Ni structure measured under different bias ranges. Sweep started from zero to positive, then reversed back to negative, and finally returned to zero in (b). (c) I-V curve of a Ni/CIPS/Ni structure device. All these curves show higher resistance ratio under positive bias. Asymmetry in (a) (Ni/CIPS/Ni structure) may be attributed to the oxidation of the bottom Ni electrode.\label{FigS3}}    
\end{figure}

\begin{figure}
    \includegraphics[width=450pt]{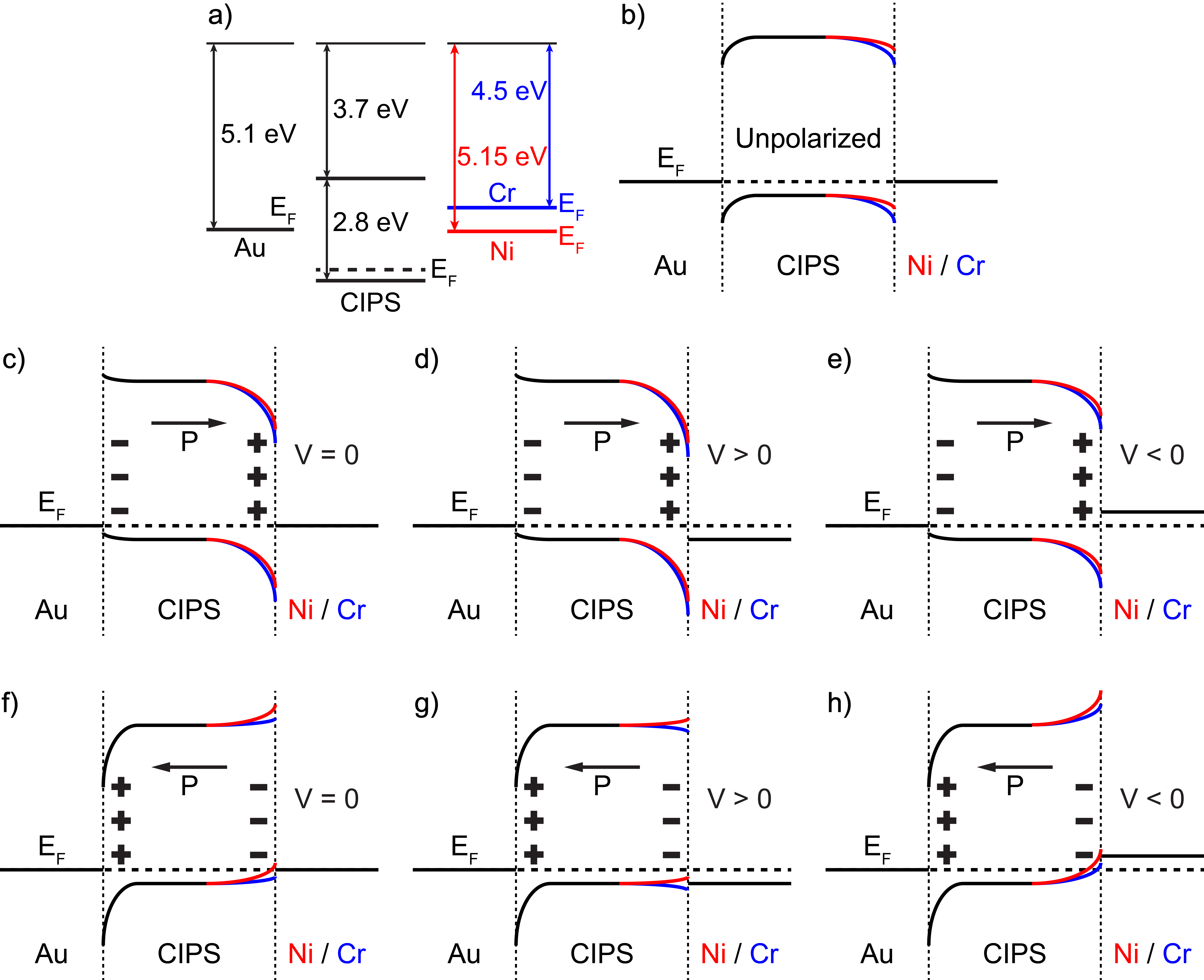}
    \caption*{FIG. S4. Band diagrams of Cr-200nm and Ni-200nm samples. a) Energy band levels of separated bottom electrode (Au), CIPS and top electrodes (Ni/Cr) (b) Band diagram when CIPS is in unpolarized state (c), (d) and (e) Band diagram when polarization points up and under zero, positive and negative bias, respectively. (f), (g) and (h) are similar to (c), (d) and (e) but with polarization pointing down.\label{FigS4}}    
\end{figure}